\documentclass[journal=apchd5,manuscript=article]{achemso}

\usepackage{multirow}
\usepackage{graphicx}% Include figure files
\usepackage{caption}
\newcommand{\sbse}{Sb\(_2\)Se\(_3\) }{}

\author{Sophie Blundell}
\affiliation{Optoelectronics Research Centre, University of Southampton, Southampton, UK}
\alsoaffiliation{School of Physics and Astronomy, University of Southampton, Southampton, UK}
\author{Thomas Radford}
\affiliation{School of Physics and Astronomy, University of Southampton, Southampton, UK}
\author{Idris A. Ajia}
\affiliation{School of Physics and Astronomy, University of Southampton, Southampton, UK}
\author{Daniel Lawson}
\affiliation{School of Physics and Astronomy, University of Southampton, Southampton, UK}
\author{Xingzhao Yan}
\affiliation{Optoelectronics Research Centre, University of Southampton, Southampton, UK}
\author{Mehdi Banakar}
\affiliation{Optoelectronics Research Centre, University of Southampton, Southampton, UK}
\author{David J. Thomson}
\affiliation{Optoelectronics Research Centre, University of Southampton, Southampton, UK}
\author{Ioannis Zeimpekis}
\affiliation{Optoelectronics Research Centre, University of Southampton, Southampton, UK}
\alsoaffiliation{Electronics and Computer Science, University of Southampton, Southampton, UK}
\author{Otto L. Muskens}
\affiliation{School of Physics and Astronomy, University of Southampton, Southampton, UK}
\email{o.muskens@soton.ac.uk}
\phone{+44 (0)23 80593911}
\title{Ultracompact programmable silicon photonics using layers of low-loss phase-change material \sbse of increasing thickness}


\begin{document}

\begin{abstract} 
High-performance programmable silicon photonic circuits are considered to be a critical part of next generation architectures for optical processing, photonic quantum circuits and neural networks. Low-loss optical phase change materials (PCMs) offer a promising route towards non-volatile free-form control of light. Here, we exploit direct-write digital patterning of waveguides using layers of the PCM \sbse with a thickness of up to 100~nm, demonstrating the ability to strongly increase the effect per pixel compared to previous implementations where much thinner PCM layers were used. We exploit the excellent refractive index matching between \sbse and silicon to achieve a low-loss hybrid platform for programmable photonics. A five-fold reduction in modulation length of a Mach-Zehnder interferometer is achieved compared to previous work using thin-film \sbse devices, decreased to 5~$\mu$m in this work. Application of the thicker PCM layers in direct-write digital programming of a  multimode interferometer (MMI) shows a three-fold reduction of the number of programmed pixels to below 10 pixels per device. The demonstrated scaling of performance with PCM layer thickness is important for establishing the optimum working range for hybrid silicon-PCM devices and holds promise for achieving ultracompact programmable photonic circuits. 
\end{abstract}

\maketitle

%%%%%%%%%%%%%%%%%%%%%%%%%%  body  %%%%%%%%%%%%%%%%%%%%%%%%%%
\section*{Introduction}

The area of photonics exploiting phase change materials (PCMs) for reconfiguring and programming functional devices has seen an enormous increase in activity, spurred by the recent availability of new materials that can be integrated in technologies such as integrated photonics and metamaterials \cite{Abdollahramezani2020TunableMaterials,PrabhatanIScience2023,ChenACSPhoton2022,Tripathi2023}. The capability to write and reset optical functionality after device fabrication is critical for a wide range of emerging applications such as quantum photonics \cite{YangNatComm2024}, optical neural networks \cite{Li2023NeuromorphicMaterials,YoungbloodNatPhoton2023}, microwave photonics \cite{MarpaungNatPhot2019}, beamforming and lidar \cite{LiLPR2022}, and optical data processing \cite{BogaertsNature2020, Zhou2022Phase-changeComputing}. Compared to fully programmable mesh structures which require a continuous energy input to maintain their state \cite{BogaertsNature2020}, non-volatile PCMs only require energy input for their initialization and reset operations, a distinct advantage compared to existing technologies \cite{Wuttig2017Phase-changeApplications,KimnjpMicrograv2024}. Programmable elements may also be used for post-fabrication diversification of photonic chips, to mitigate the lengthy cycle of application-specific silicon photonics device design and fabrication \cite{BogaertsIEEE2020}. 

Applications of non-volatile tuneable and reconfigurable photonics depend critically on the development of new optical PCMs that can provide control over the optical phase of light without introducing large losses \cite{PrabhatanIScience2023}, while maintaining a high endurance for repeated cycling between the material states \cite{Martin-MonierOMEX2022}. Following the introduction of the low-loss PCM antimony selenide (\sbse) for silicon photonics \cite{DelaneyAdvFMat2020}, devices exploiting thin \sbse{} films on top of photonic waveguides have seen a highly successful development in recent years by an increasing number of researchers worldwide, which has included demonstrations of direct-write optical programming \cite{DelaneySciAdv2021,WuSciAdv2024} as well as electrical switching using integrated pn-junctions \cite{RiosPhotonix2021,FangNanoLett2024} and graphene microheaters \cite{FangNatNanotech2022}. Most demonstrations shown to date have used a very thin layer of PCM material of around 20~nm thickness. The dependence of performance of programmable silicon photonics on the PCM layer thickness has remained largely unexplored.

The low-loss optical PCM \sbse{} is of particular interest for silicon photonics because of its refractive index which is closely matched to silicon, allowing for seamless integration, furthermore it exhibits a sizeable switching contrast between the amorphous and crystalline states of $\Delta n = 0.77$ and shows very low intrinsic material losses around the telecommunications band at 1550~nm wavelength \cite{DelaneyAdvFMat2020}. A reversible change in optical phase exceeding $\pi$ was demonstrated through direct optical writing of a 23-nm thick slab of \sbse{} deposited on top of a 220~nm SOI rib waveguide located in one of the arms of an asymmetric Mach-Zehnder Interferometer (MZI) device \cite{DelaneySciAdv2021}. With this low thickness of \sbse{}, the optical phase shift was found to be around $0.04\pi$ per $\mu$m, resulting in a device length $L_\pi$ of around 25~$\mu$m. Subsequent other works reported similar magnitudes of the effect exploiting electrical switching \cite{RiosPhotonix2021,FangNatNanotech2022,FangNanoLett2024}. While these very thin films of the low-loss PCM \sbse{} offer good performance for optical phase control of silicon photonic devices, increasing the optical phase shift per device length would be of interest to achieve more compact device geometries. Thicker PCM layers could provide a shorter device length by increasing the switching contrast of the mode index between the material states, however this has to be traded off against increased losses by absorption, scattering and multi-mode contributions. 

In this work, we investigate the role of the thickness of low-loss PCM \sbse{} on the performance of programmable silicon-on-insulator (SOI) photonic devices. We exploit the excellent matching of the refractive index of \sbse{} to silicon to ensure good mode overlap and low insertion losses in the amorphous phase. Integration of the low-loss optical PCM \sbse in a silicon photonic device provides a method for controlling the optical phase  without introducing very large transmission losses. The main figure of merit for hybrid silicon-PCM devices is the induced optical phase shift between the switched states, $\Delta \phi$, normalized against the device loss, $\alpha$, caused by scattering and/or absorption in the waveguide. In the devices under study, $\alpha$ is governed primarily by losses in the crystalline state.

The first part of this work is aimed at determining this figure of merit of hybrid silicon-\sbse devices with increasing \sbse thickness. The device design and fabrication is introduced, followed by optical studies of device losses $\alpha$ in straight waveguides. The induced optical phase shift $\Delta \phi$ is studied in Mach-Zehnder interferometer (MZI) devices to arrive at the device switching figure of merit $\Delta \phi/\alpha$. The second part of the studies investigates the capability of thicker PCM layers in digital patterning of multimode interference devices (MMIs). Digital patterning of weak perturbations, as described by us in previous works\cite{Bruck2016,DelaneySciAdv2021,Dinsdale2021} is distinct from topological inverse design which relies on precise shaping of complex patterns with ultrafine features \cite{Wu2024, nikkhah2024}. Digital patterning is typically achieved over a much coarser grid of perturbation coordinates, making use of multiple scattering and mode mixing at the individual perturbations \cite{Bruck2016}. Here, we investigate how the number of pixels required to control a multimode device scales with the PCM thickness. An optimal working point in the number of pixels required to switch the device, achieving a high level of control over the output state whilst minimizing programming effort, is of interest for achieving a platform for digital programmable photonics \cite{DelaneySciAdv2021, Dinsdale2021}.

\section*{Results and Discussion}\label{results}

\subsection*{Design and fabrication of hybrid Silicon-\sbse devices}

Silicon photonic devices were fabricated using a 220 nm silicon on insulator (SOI) platform \cite{Littlejohns2020}, as explained in the Methods section. Photonic rib waveguides of 120 nm thickness were covered with patches of the low-loss PCM \sbse. A 50-nm thin SiO$_2$ cladding was used as a protection layer on top of the \sbse PCM.

\begin{figure*}[tbhp!]
    \centering
    \includegraphics[width=0.95\textwidth]{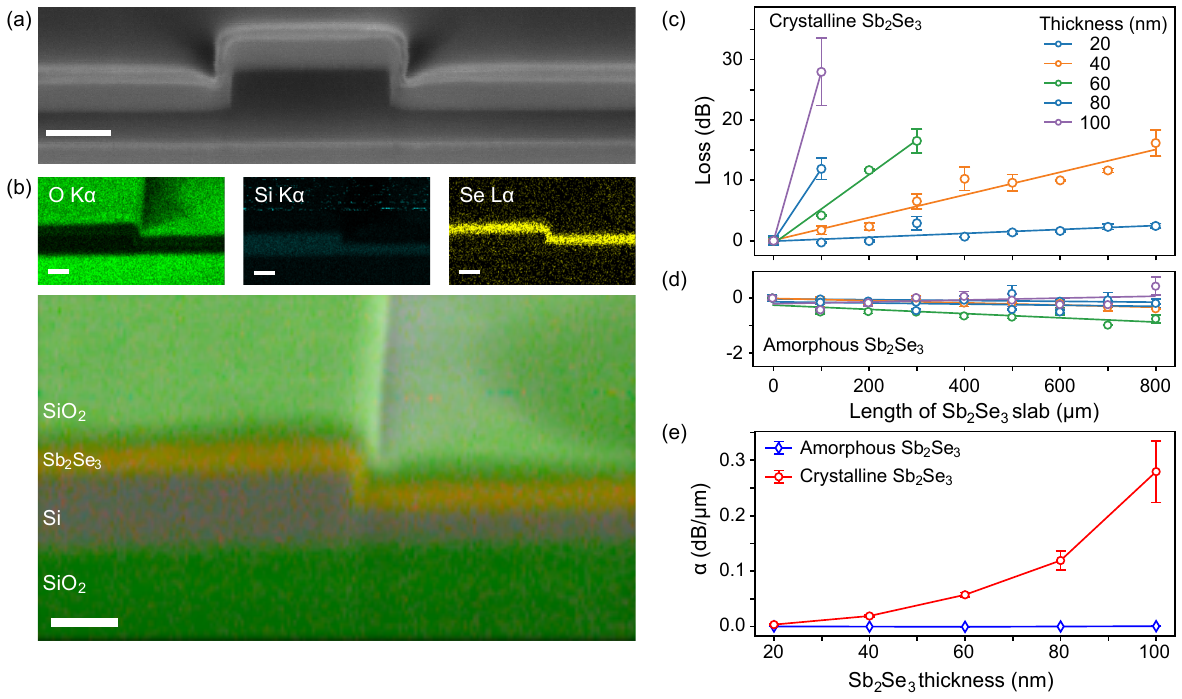}
    \caption{(a) Cross sectional SEM image of silicon waveguide with \sbse layer of 100~nm thickness and 20~nm SiO$_2$ cladding. (b) Energy dispersive spectroscopy (EDS) analysis of MMI device showing oxygen (O K$\alpha$), silicon (Si K$\alpha$) and selenium (Se L$\alpha$) signatures in separate panels and in SEM overlay (large panel). All scale bars in (a,b) are 200~nm. 
    (c,d) Measured insertion loss of straight waveguides with varying lengths of Sb\textsubscript{2}Se\textsubscript{3} embedded for five different deposition thickness, for crystalline state (c) and amorphous state (d), normalized to straight waveguide without PCM. (e) Propagation loss $\alpha$ in dB/$\mu$m against \sbse layer thickness, for amorphous state (diamond, blue) and crystalline state (circle, red).}
    \label{fig:WGloss}
\end{figure*}

Figure~\ref{fig:WGloss}(a) shows a Scanning Electron Microscope (SEM) cross-section of  a silicon rib waveguide covered with an \sbse layer of 100~nm thickness. The SEM allows identification of different layers in the stack through their contrast in the backscattered electron image, the \sbse layer being the light gray region on top of the darker SOI waveguide. More specific elemental information is obtained from Energy Dispersive Spectroscopy (EDS) at 2~kV, as shown in Figure~\ref{fig:WGloss}(b) for a region of a MMI device. The EDS map clearly identifies the \sbse layer through the Se L$\alpha$ contribution. We find good agreement between the designed and fabricated thickness of the \sbse layer. The aim of an Ar-plasma pre-treatment in the process flow was to achieve a partial embedding of the PCM into the waveguide, however the cross section reveals that the Ar-plasma only marginally affected the Si thickness and the \sbse layer sits on top of the waveguide. A discussion of the device geometry, with numerical simulations of the effective index contrast for different embedding depths, is presented in more detail in the Supporting Information Section S1. The difference between embedded and non-embedded PCM in the optical switching response is found to be modest and equivalent to a variation of around 20~nm of the PCM thickness itself.

\subsection*{Characterization of straight waveguide losses}

The fabricated chips feature a series of straight waveguide (SWG) devices functionalized with slabs of \sbse{} PCM ranging from 0~$\mu$m to 800~$\mu$m in length. Insertion loss measurements were performed at 1550~nm wavelength to determine the waveguide loss versus length, both for as-deposited amorphous samples as well as hot-plate crystallized \sbse (see Methods). Results are presented in Fig.~\ref{fig:WGloss}c and d for the crystalline and amorphous \sbse respectively. All results for PCM cladded devices were normalized to a straight SOI rib waveguide of the same total length, represented by the data point at zero length of \sbse slab. Error bars were obtained from the measured variation over three different waveguides on the same chip, for each length of the \sbse patch.

Figure~\ref{fig:WGloss}c shows that losses increase exponentially with the length of \sbse, following the Beer-Lambert law, which on a dB scale shows a linear trend with the slope equal to the loss coefficient $\alpha$ in dB/$\mu$m. In the amorphous state, losses stay very low for increasing thickness and length of \sbse and in fact the waveguide loss is reduced compared to the reference waveguide without PCM, resulting in a negative loss coefficient $\alpha$. The reduced loss can be attributed to the increase of the waveguide cross section, resulting in reduced surface scattering compared to the original SOI device.

Waveguide loss coefficients $\alpha$ were extracted using linear fits to the experimental data, as shown by the lines in Fig.~\ref{fig:WGloss}c, d, and resulting values are presented in Fig.~\ref{fig:WGloss}e against \sbse thickness, both for the crystalline (red dots) and amorphous (blue diamonds) states. Corresponding values for $\alpha$ in the crystalline state are tabulated in Table~\ref{table1}. Figure~\ref{fig:WGloss}e clearly shows the rapidly increasing loss for thicker \sbse layers, with a factor of two higher loss factor for each 20~nm increase in thickness, and reaching up to 0.28~dB/$\mu$m for the 100~nm \sbse devices. 

\subsection*{Characterization of switching-induced optical phase shift in silicon-\sbse MZIs}

\begin{figure*}[tbh]
    \centering
    \includegraphics[width=0.95\textwidth]{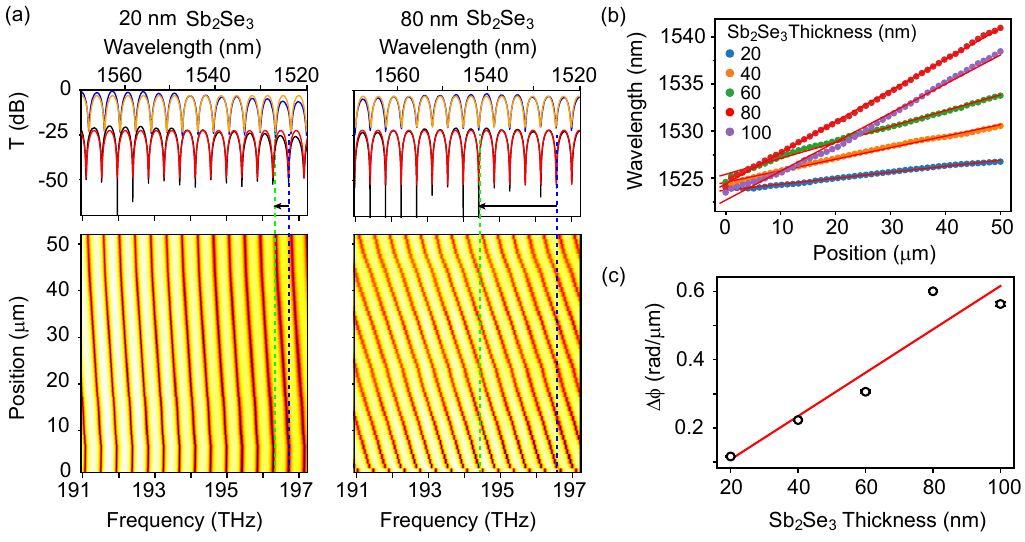}
    \caption{(a) Spectra of MZIs before and after crystallization (top panels) and maps taken after every 1$\mu$m of crystallization (bottom panels) by direct laser writing over 50$\mu$m of \sbse{}, for devices with \sbse thickness of 20~nm and 80~nm. Black/blue curves in top panel: experimental spectra, red/orange curves: model fits. Vertical dashed lines indicate shift of selected mode with initial wavelength around 1524~nm. (b) Wavelength position extracted from fits to experimental spectra for MZI devices with \sbse thickness between 20~nm and 100~nm. Red lines: linear fits to data. (c) Extracted values of slope $\Delta \phi$ plotted against \sbse thickness. Red line: linear fit.}
    \label{fig:crysspec}
\end{figure*}

The induced optical phase shift was characterized using the spectral response of asymmetric MZI devices containing a patch of \sbse in one of the arms. Measurements were taken of spectra between 1520~nm and 1570~nm while the \sbse was incrementally crystallized using direct laser writing in 1$\mu$m steps, using the setup described in the Methods section. Results are shown in Fig.~\ref{fig:crysspec}(a) for MZI devices with the 20~nm and 80~nm thick \sbse layers, results for the other layers are presented in the Supporting Information. The top panels show the experimental transmission spectra of the device before (black) and after (blue) switching. 

Fits using the typical periodic MZI comb function are also presented by the red and orange curves, respectively, allowing to extract precise values for the free spectra range, wavelength shift and contrast taking into account the contribution from all peaks in the spectrum. The fitting procedure is discussed in more detail in the Supporting Information.

Full experimental maps of the spectra versus switching length along the \sbse slab are shown in the bottom panels in Fig.~\ref{fig:crysspec}(a), showing the results of 51 individual spectra measured as the direct write laser was scanned along 50~$\mu$m of the PCM patch.

For each device, we selected a transmission minimum around 1525~nm (196.6~THz) and followed the position of this peak along the switching position, using the fitted spectra to provide the most accurate results. The wavelength shift versus position is shown in Fig.~\ref{fig:crysspec}(b) for all five thickness values of the \sbse layer. For all samples, we see a wavelength shift of this feature proportional to the length of the switched region, which could be fitted (lines) to obtain a wavelength shift $\Delta \lambda$ per unit of length. The wavelength shift $\Delta \lambda$ is converted to a phase shift $\Delta \phi$ by considering the fitted free spectral range (FSR) of the MZI spectra, which is equivalent to $2\pi$ radians of phase shift
\begin{equation}
    \Delta\phi=\frac{\Delta\lambda}{\lambda_{\rm FSR}}2\pi.
\end{equation}

Figure~\ref{fig:crysspec}(c) presents the resulting values of $\Delta \phi$ for the set of devices with different \sbse layer thickness. Similar results were obtained for two other sets of devices, containing a 10~nm thin SiO$_2$ buffer layer between the silicon waveguide and the PCM, as presented in the Supporting Information. Altogether these results support a trend showing a linear increase of the induced optical phase shift $\Delta \phi$ with \sbse thickness in the range from $20-100$~nm. In terms of the FSR itself, fitting of the frequency transfer function of the MZI, presented in the Supporting Information Figure S7, shows a $\sim 1.2$\% reduction of the FSR frequency after switching of a 50~$\mu$m long slab of the 100~nm thick \sbse layer. The mode at 1525~nm wavelength corresponds to the 475th free spectral range (FSR) in the frequency spectrum, the FSR being 0.41385~Thz, or 3.208~nm at 1525~nm. 

\subsection*{Device Figure of Merit}

To quantify the overall performance taking into account both optical phase shift and losses, we use a device figure of merit (FOM) introduced previously in Ref.~\cite{DelaneyAdvFMat2020}, defined by the ratio of the optical phase shift over the device loss
\begin{equation}\label{eqn1}
    FOM=\frac{\Delta \phi}{\alpha}
\end{equation}
Compared to the conventional materials FOM $\Delta n/\Delta k$, which only takes into account bulk properties, the device FOM provides a more useful value of the performance in a waveguide geometry. Given the low intrinsic values for the material losses in antimony-based PCMs \cite{DelaneyAdvFMat2020}, the materials FOM diverges in the near-infrared \cite{Abdollahramezani2020TunableMaterials}, requiring the use of a more realistic performance figure such as defined in Eq.~\ref{eqn1}. The device FOM allows us to consider not just the change in refractive index between the two phases of the materials, but to take into account the inherent losses of the material and structure to gain a full appreciation of how useful this material is in this configuration.

For the FOM we use the propagation loss of the crystalline phase of \sbse, as this is typically much higher than in the amorphous phase and is the performance limiting value. Performing this calculation for our range of thicknesses, we arrive at results presented in Table~\ref{table1}. The device FOM reaches a value around 35 for the thinnest \sbse layers under study, consistent with earlier reported findings \cite{DelaneyAdvFMat2020}. Increasing the thickness of the PCM results in a sharp drop of the FOM because of the additional waveguide losses which increase much more rapidly with thickness than the induced phase shift. However, as long as losses of the order of 1~dB can be tolerated, there is much to gain in terms of device length by increasing the PCM thickness. The reduction in device length can be appreciated from the calculated values for $L_\pi$ in Table~\ref{table1}. A fivefold reduction of $L_\pi$ from 27.1~$\mu$m down to 5.2~$\mu$m is obtained by increasing the \sbse thickness from 20~nm to 80~nm, at the cost of $0.62$~dB insertion loss over the device length compared to 0.08~dB for the 20~nm thin film. Table \ref{table1} thus demonstrates the trade-off between the benefit of increasing the thickness of \sbse, increased modulation, and the disadvantage, increased propagation losses.

\begin{table*}[h!]
\begin{tabular}{|p{2.5cm}|rl|rl|rl|r|}
\hline
\sbse{} thickness(nm) & \multicolumn{2}{|c|}{$\alpha$ (dB/ \(\mu\)m)} & \multicolumn{2}{|c|}{$\Delta \phi$ (rad/ \(\mu\)m)} & \multicolumn{2}{|c|}{FOM (rad/dB)}  & $L_\pi$ (\(\mu\)m)\\
\hline
20&0.003&$\pm$ 0.001 & 0.116&$\pm$ 0.002 & 35.9&$\pm$ 0.50 & 27.1\\
40&0.019&$\pm$ 0.002 & 0.223&$\pm$ 0.002 & 11.8&$\pm$ 0.10 & 14.1\\
60&0.057&$\pm$ 0.004 & 0.306&$\pm$ 0.004 & 5.35&$\pm$ 0.07 & 10.3\\
80&0.120&$\pm$ 0.020 & 0.600&$\pm$ 0.002 & 5.04&$\pm$ 0.02 & 5.2\\
100&0.280&$\pm$ 0.060 & 0.563&$\pm$ 0.007 & 2.01&$\pm$ 0.04 & 5.6\\
\hline
\end{tabular}
\caption{Experimental propagation loss \(\alpha\),  optical phase shift $\Delta \phi$ and calculated figure of merit (FOM) for SOI waveguides with different thickness of embedded \sbse{}.}
\label{table1}
\end{table*}

\subsection*{Digital patterning of MMIs using direct laser writing}

Having established the effect of increased \sbse layer thickness on the induced optical phase shift in a MZI configuration, we proceed our investigation of digital pixel patterning of MMIs for programmable multi-port photonic circuits. Experiments were done starting from simulated pixel patterns, which were obtained using a  forward iterative optimization similar to  \cite{Bruck2016,Dinsdale2021}. To take into account uncertainty in the experimental device cross section, the transmission for the patterns was re-simulated for different embedding depths of the PCM into the waveguide.

Experimentally, transmission of both output ports was measured simultaneously using a custom built dual-fibre probe \cite{DelaneySciAdv2021}. Intensity at both outputs was measured during the direct laser writing and was recorded for each pixel, allowing to extract total device throughput and transmission contrast between the output ports. In our studies we considered both amorphous pixels on a pre-crystallized \sbse device, similar to previous works \cite{DelaneySciAdv2021}, as well as crystalline pixels on an as-grown amorphous \sbse layer.

\subsubsection*{Amorphous pixels on crystalline MMI}

\begin{figure*}[tbh]
    \centering
    \includegraphics[width=1.0\textwidth]{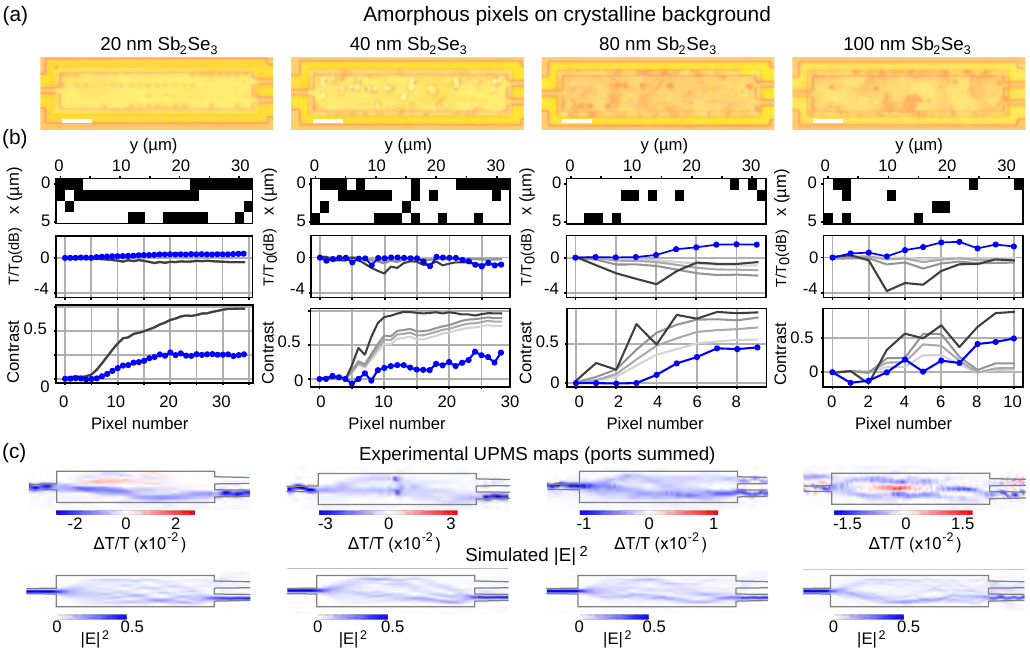}
    \caption{(a) Optical microscopy images of patterned MMIs for \sbse samples of 20, 40, 80, 100~nm thickness, For amorphous pixels written on a pre-crystallised PCM using direct laser writing. (b) Designed perturbation maps showing switched pixels (top). Total transmission over both output ports $T=T_{\rm btm}+T_{\rm top}$, normalised to transmission before switching, $T_0$ (middle), and contrast between top and bottom output ports $(T_{\rm btm}-T_{\rm top})/T$ (bottom), shown versus switched pixel number. Experimental data (blue dots) and simulations from 0\%, 25\%, 50\% to 100\% etch depth (light to dark grey lines). (c) Ultrafast photomodulation spectroscopy (UPMS) maps of total port transmission, $\Delta T / T$ (top) and calculated near-field intensity maps $|E|^2$ over the MMI devices.}
    \label{fig:pixpat_crybg}
\end{figure*}

Figure~\ref{fig:pixpat_crybg} shows results for four selected MMIs with \sbse thickness ranging between 20~nm and 100~nm. The samples were pre-crystallized using a hotplate and amorphous pixels patterns were programmed by scanning of a direct-write laser over the area. Optical microscopy images are presented in Fig.~\ref{fig:pixpat_crybg}(a) and the target design patterns are shown in (b). Fig.~\ref{fig:pixpat_crybg}(b) also shows the overall device throughput $T$ summed over both bottom ($T_{\rm btm}$) and top ($T_{\rm top}$) output ports, normalized to the initial value before patterning $T_0$, as well as the port contrast defined as $(T_{\rm btm}-T_{\rm top})/T_0$. Experimental results (blue dots) are compared with simulations for a range of embedding depths from 0\% (light grey), 25\%, 50\%, and 100\% (dark grey) of the design geometry. 

Application of the pattern results in an increased transmission of the bottom output. A lower contrast is seen for the experiments compared to the simulations, even when accounting for deviations in etch depth. We attribute the different to the difference in pixel size between the experiment of Fig.~\ref{fig:pixpat_crybg} (a) to the design of (b), which resulted in weaker perturbation strength in the experiments than in the model. 

Furthermore, we see that much less pixels are needed to induce the same change for the thicker \sbse layers. For the device throughput, we see that simulations show the opposite effect than experiments, where an improvement of transmission is obtained by pixelating. This can be explained by the fact that, in our simulations, we do not take into account the scattering-induced losses in the crystalline state. The simulation therefore overestimates the throughput of the unperturbed devices. 

\subsubsection*{Ultrafast photomodulation spectroscopy}
Whereas the port transmissions provide important information about the effect of digital patterning of the device, additional information can be gained by looking at the local field distribution inside the device. In this work, we have investigated for the first time the flow of light inside MMIs with PCMs using ultrafast photomodulation spectroscopy (UPMS), a technique capable of mapping the flow of light in space and time \cite{Bruck2014Device-levelSpectroscopy}.  In UPMS an ultrafast pump laser focus is scanned across the device, while the transmission through the waveguide is monitored using a pulsed probe laser \cite{Vynck2018UltrafastDevices}. Previously we have obtained results for waveguides patterned by etching of holes into the silicon \cite{Dinsdale2021}. Figure~\ref{fig:pixpat_crybg}(c) shows experimental UPMS maps taken of the patterned devices, where the differential transmission $\Delta T/T$ is shown summed over both output ports. A detailed model explaining the UPMS response can be found in \cite{Vynck2018UltrafastDevices}. In short, a negative $\Delta T/T$ response indicates a reduced port transmission while a positive value means an increased port transmission in the presence of the ultrafast pump laser. 

For a device with high throughput, the summed-port response is predicted to follow closely the local near-field intensity profile in the device \cite{Vynck2018UltrafastDevices}. We compare our results directly to the calculated near-field intensity $|E|^2$, presented in \ref{fig:pixpat_crybg}(c), bottom panels. A Gaussian smoothing filter was applied to take into account the experimental resolution, blurring out some of the finer features but retaining the light flow profiles in the different patterned MMIs. The maps show that different pixel patterns result in varying flow profiles as is the case for the experimental UPMS maps. In the absence of losses, the port-summed UPMS maps should ideally show only a perturbation induced loss and hence an overall blue color. For the 40~nm and 80~nm \sbse layers, this is indeed observed and the demonstrated flow patters are in good overall agreement between the experimental and simulated maps. Branched flows are typically observed in weakly perturbed waveguides \cite{Brandstotter2019} and are indicative of the regime of weak forward scattering.

The presence of positive $\Delta T/T$ response in the maps taken for the 20~nm and 100~nm \sbse MMIs in Fig.~\ref{fig:pixpat_crybg}(c) is indicative of positions on these MMIs where an additional perturbation increases the throughput. This indicates that the device has additional loss channels related to a mismatch between the designed and experimentally applied pixel patterns. For example, for the 100~nm \sbse device, the calculated near-field map shows that the light splits up into two branched flows at the top and bottom of the device. This branching into a top and bottom flow is also seen experimentally. Additionally we see that placing the experimental perturbation in the middle of the MMI improves the throughput, indicating that some of the light did not follow the designed flow pattern, but the extra pump-induced perturbation acts to reroute some of the misdirected light back to the outputs. Similarly, for the 20~nm \sbse it appears that the long row of pixels at the top of the device was insufficient in rerouting the light, placing the perturbation in this area also improves that routing of light toward the MMI outputs. 

\subsubsection*{Crystalline pixels on as-grown amorphous MMI}

\begin{figure*}[tbh]
    \centering
    \includegraphics[width=1.0\textwidth]{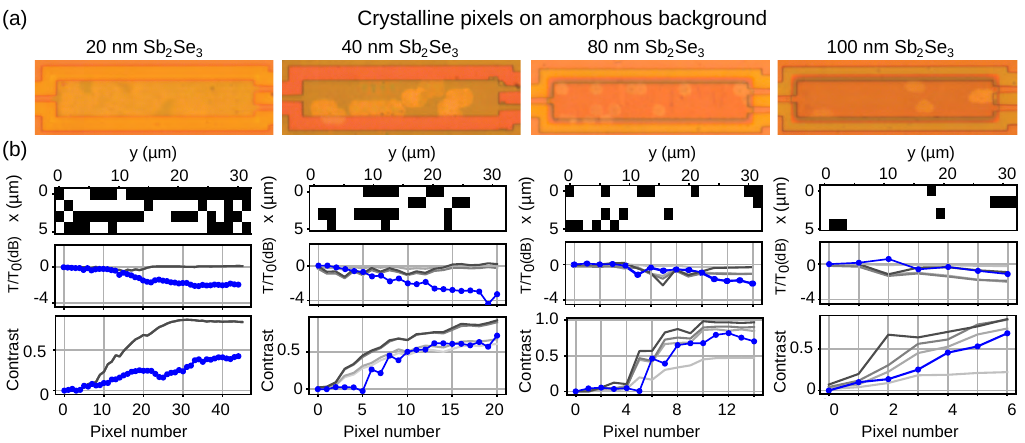}
    \caption{(a) Optical microscopy images of patterned MMIs for \sbse samples of 20, 40, 80, 100~nm thickness, For crystalline pixels written on as-grown amorphous PCM using direct laser writing. (b) Designed perturbation maps showing switched pixels (top). Total transmission over both output ports $T=T_{\rm btm}+T_{\rm top}$, normalised to transmission before switching, $T_0$ (middle), and contrast between top and bottom output ports $(T_{\rm btm}-T_{\rm top})/T$ (bottom), shown versus switched pixel number. Experimental data (blue dots) and simulations from 0\%, 25\%, 50\% to 100\% etch depth (light to dark grey lines).}
    \label{fig:pixpat_amobg}
\end{figure*}

The method of direct laser writing of amorphous pixels on a crystalline background, such as shown in \ref{fig:pixpat_crybg}, results in small pixels with well-defined edges. This can be understood as the time scales associated with vitrification are of the order of tens of nanoseconds, much shorter than times for lateral heat diffusion. However, this approach requires the entire PCM layer to be crystallized, which means that most of the PCM is in the state where losses are increased compared to the original device.

Next, we investigate the possibity of starting from the as-grown, amorphous device and induce patterning by selective crystallisation of pixels. Crystallization offers limited spatial control because of lateral heat diffusion and crystallization dynamics, generally resulting in larger pixel size compared to amorphous pixels resulting from rapid vitrification. Here we explore this route experimentally as shown in Fig.~\ref{fig:pixpat_amobg}. Patterns were designed numerically before the experiment following the same methodology as before but using crystalline pixels on an amorphous PCM background. Direct laser writing in this configuration results in much larger crystallized regions, as can be seen in Fig.~\ref{fig:pixpat_amobg}(a) for all of the \sbse layer thickness values, which is expected from the different time scale allowing for significant lateral heat diffusion and crystal growth. In the 20~nm \sbse layer, where patterns exceeding 40 pixels were required, the individual laser-written spots are seen to fuse into a semi-continuous crystalline region with very little remaining structure. For the other three devices, distinct pixels can still be identified.

The throughput and switching contrast of the devices are shown in Fig.~\ref{fig:pixpat_amobg}(b). The devices show a good switching contrast owing to the larger pixel size than in the amorphous case. However the large pixel size limits the number of pixels that can be written for the 20~nm \sbse device, where the contrast is reduced due to the loss of pattern structure due to pixel fusion. As can be seen in the $T/T_0$ trends, the writing of crystalline pixels on an amorphous PCM background lowers the total transmission by up to 4~dB compared to the original device, this is in agreement with the expectation based on additional losses in the crystalline state. Simulations underestimate this device loss, again because of the absence of scattering by polycrystalline domains. The 80~nm and 100~nm \sbse devices in this set of devices show the best agreement with simulations, as the pixel size is very well matched and the number of pixels is sufficiently small to allow for each pixel to act effectively on its own. UPMS measurements were attempted on this set of devices using the same configuration as in Fig.~\ref{fig:pixpat_crybg}(c), but the effectiveness of the technique was strongly reduced by a very low photomodulation response for the \sbse in its amorphous state. Further work is needed to more systematically explore this difference in response which goes beyond the scope of the current paper.

\begin{figure*}[tbh]
    \centering
    \includegraphics[width=0.75\textwidth]{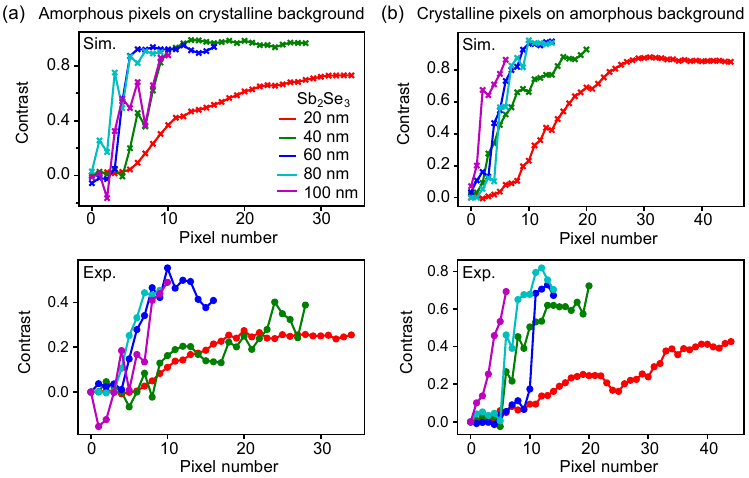}
    \caption{Contrast between bottom and top output ports against pixel number for five patterned MMIs with different \sbse thickness between 20~nm and 100~nm, for amorphous pixels on crystalline background (a) and crystalline pixels on as-grown amorphous background (b). Top graphs, simulation results for design patterns; bottom graphs, experimental results.}
    \label{fig:allcontrast}
\end{figure*}

Combining our results for the amorphous and crystalline pixel patterns, Fig.~\ref{fig:allcontrast} summarizes the port contrast against pixel number for the investigated devices. This set includes the 60~nm \sbse layer device which data are included in the Supporting Information. The results support the overall conclusion that thicker layers of the low-loss PCM \sbse allow for increased perturbation strength, reducing the number of pixels needed to achieve a given splitting ratio using direct write digital patterning of a MMI device. Results for the amorphous pixels were somewhat lower mainly because of the smaller pixel size, whereas crystalline pixels provided an inherently much larger pixel size, which posed challenges for defining intricate designs with many pixels, but for the thicker \sbse layers was otherwise very effective in routing light using only a small number of pixels. The results for digital patterning of MMIs are in agreement with the larger induced phase shift observed in our MZI calibration, which formed the basis of our numerical pattern designs.

\section*{Conclusion}\label{conc}
In conclusion, we have investigated the dependence of the optical switching efficiency of layers of different thickness of the low-loss phase change material \sbse when integrated onto a standard 220-nm silicon photonics platform. Propagation loss measurements on straight waveguides were combined with precise measurements of the induced optical phase shift in a Mach-Zehnder interferometer configuration, to arrive at a device figure of merit for each layer thickness. A large, five-fold increase of the induced phase shift was observed by increasing the PCM thickness from 20~nm to 80~nm, resulting in a device length reduction $L_\pi$ from $27.1$~$\mu$m down to 5.2~$\mu$m. This strong reduction opens possibilities for new types of ultracompact photonic devices. The reduced footprint has to be traded off against increased optical losses, increasing to an insertion loss of around 0.62~dB for the 80~nm thick \sbse layer. The calibration of optical phase shift was subsequently used to design new types of programmable photonic routers based on direct write digital patterning of a multimode interference device. The application of pixel patterns was shown to result in branched flows of light streaming from the input to the selected output, which was experimentally visualized using  ultrafast photomodulation spectroscopy mapping. The increased thickness of \sbse resulted in a reduction of the required pixels from larger than 30 to less than 10 per patterned device, which significantly reduces the complexity of the scheme. Amorphous pixels on a crystalline background were shown to offer precise resolution and control over the pixels, however at the cost of increased losses in the pre-crystallized background. In comparison crystalline pixels on an amorphous background were demonstrated as an alternative showing larger pixel size with reduced control, but stronger perturbations per pixel which worked particularly well for thicker \sbse layers. The demonstrated new capability has relevance in post-fabrication tuning of silicon photonic devices and may hold promise for achieving reconfigurable and freeform programmable devices for optical processor technology, photonic AI hardware, and quantum computing.

\begin{acknowledgement}

The authors thank M. Ebert, G. Mourkioti and N. Zhelev for technical support on cross-sectional SEM-EDX imaging. This work was supported financially by EPSRC through grant EP/M015130/1. Silicon photonic waveguides were manufactured through the UK Cornerstone open access Silicon Photonics rapid prototyping foundry through EPSRC grant EP/L021129/1. D.J.T. acknowledges funding from the Royal Society for his University Research Fellowship.

\end{acknowledgement}

%%%%%%%%%%%%%%%%%%%%%%%%%%%%%%%%%%%%%%%%%%%%%%%%%%%%%%%%%%%%%%%%%%%%%
%% The same is true for Supporting Information, which should use the
%% suppinfo environment.
%%%%%%%%%%%%%%%%%%%%%%%%%%%%%%%%%%%%%%%%%%%%%%%%%%%%%%%%%%%%%%%%%%%%%
\begin{suppinfo}

Supporting simulation results, experimental procedures, description of model fits and additional experimental data.

\end{suppinfo}

\section{Methods}
\textbf{Waveguide fabrication.}
Silicon-on-insulator (SOI) waveguide devices were fabricated by deep-UV lithography on 200~mm diameter SOI wafers using the Cornerstone silicon photonics foundry at the University of Southampton, UK. In a first exposure, photonic rib waveguides of 120~nm in height were fabricated onto the 220 nm SOI platform using an etching step. A second deep-UV exposure was used to define the open windows in a photoresist at selected locations on the wafer to allow subsequent etching and PCM deposition. The wafer was cleaved into chips.

Individual chips were further processed by loading into a sputter tool (AJA Orion). A first Ar-ion etch was done using 30 W of RF bias in Argon in order to remove any residual resist. Subsequently, the \sbse and cladding stack was deposited without breaking the vacuum. For results presented in the main paper, the \sbse{} was deposited directly onto the silicon waveguide. A second set of samples was produced including a 20-nm thin SiO$_2$ buffer layer between the silicon waveguide and the \sbse. The \sbse{} layer was sputtered from a stoichiometric \sbse{} target (Testbourne) with a thickness proportional to sputtering time. A 20-nm thin SiO$_2$:ZnS (80:20 ratio) dielectric was deposited as a thin protective layer. The photoresist was then removed using NMP, Acetone, and IPA. The final step was to clad the sample, which was performed by sputtering 100~nm of SiO$_2$:ZnS over the entire chip.%

\textbf{Crystallization using thermal annealing.}
Thermal annealing was performed for selected devices using a hot plate.
To crystallize the \sbse{} integrated into photonic chips, they were heated to 190$^{\circ}$C for 10 minutes, after which they were inspected using an optical microscope.

\textbf{Optical spectroscopy}
Insertion loss spectra of as-grown amorphous and pre-crystallized samples were done using a standard insertion loss measurement system. Studies involving direct laser writing with sequential spectroscopy were done using a customized setup including a direct write diode laser (Vortran stradus) at 638~nm wavelength and 150~mW peak power.
Short digital trigger pulses were generated using a Berkeley Nucleonics pulse generator in order to produce a variety of pulse lengths and powers. 
A 50X, 0.5 numerical aperture objective was used (Mitutoyo) and the objective position was scanned using a closed-loop piezo nanopositioning system. 

A broadband swept tunable laser source (Keysight N7778C) was used for sequential scans of MZIs in combination with a multi-port power meter (Keysight N7744A). For the two-port output detection, a custom-built two-fibre arm was used to simultaneously collect both outputs. The setup was controlled using a computer interface (Labview).

\textbf{Ultrafast Photomodulation Spectroscopy}
For ultrafast mapping we use a setup similar to that reported in Refs. ~\cite{Bruck2014Device-levelSpectroscopy, Vynck2018UltrafastDevices}. An ultrafast fibre laser (Menlo) at 1550~nm wavelength was used as the probe at 100 MHz repetition rate. Part of its output was quadrupled to 390~nm using two second harmonic crystals resulting in several mW optical power at the UV wavelength. The UV output was amplitude modulated at 10~MHz using an acousto-optic crystal and was focused on the top of the device under test using a 0.5 NA objective (Mitutoyo) with closed-loop nanopositioning system for scanning (Smaract). Light was detected using an APD (Thorlabs) and a lock-in amplifier (Zurich Instruments). 
 
\bibliography{references}

\end{document}